
\baselineskip=20pt
\magnification=\magstep1

\def\C{{\bf C}}

\def\Z{{\bf Z}}

\def\la{\lambda}
\def\La{\Lambda}

\def\V1{V^{(1)}}


\def\End{\hbox{\rm End}}
\def\id{\hbox{\rm id}}

\def\tr{\hbox{\rm tr}}


\def\slt{\goth{sl}_2}
\def\slth{\widehat{\goth{sl}}_2\hskip 2pt}

\def\goto#1{{\buildrel #1 \over \longrightarrow}}
\def\br#1{\langle #1 \rangle}
\def\bra#1{\langle #1 |}
\def\ket#1{|#1\rangle}

\def\Rc{\check R}
\def\wt{\hbox{\rm wt}\,}

%
\font\germ=cmmi10
\def\goth#1{\hbox{\germ #1}}
%
\def\sectiontitle#1\par{\vskip0pt plus.1\vsize\penalty-250
 \vskip0pt plus-.1\vsize\bigskip\vskip\parskip
 \message{#1}\leftline{\bf#1}\nobreak\vglue 5pt}
\def\qed{\hbox{${\vcenter{\vbox{
    \hrule height 0.4pt\hbox{\vrule width 0.4pt height 6pt
    \kern5pt\vrule width 0.4pt}\hrule height 0.4pt}}}$}}
\font\bxit=cmbxti10
\def\subsec(#1|#2){\medskip\noindent{\bxit #1}\hskip8pt{\bxit #2}\quad}
\def\eq#1\endeq
{$$\eqalignno{#1}$$}
\def\leq#1\endeq
{$$\leqalignno{#1}$$}
%
%
\def\Proof{\noindent {\sl Proof.\quad}}

%
%

%
%
\def\Definition#1.#2{\smallskip\noindent {\sl Definition #1.#2\quad}}
%
%
%
%
%
%
\def\abstract#1\endabstract{
\bigskip
\itemitem{{}}
{\bf Abstract.}
\quad
#1
\bigskip
}
%
%
%
\def\sec(#1){Sect.\hskip2pt#1}


\def\u{U_q\bigl(\slth \bigr)}
\def\Up{U'_q\bigl(\slth\bigr)}

\def\VO(#1,#2){{\tilde \Phi}^{#1}_{#2}}
\def\h(#1){|u_{#1}\rangle}

\def\e{\hbox{\rm e}}

\def\ov#1#2{{#1 \over #2}}
\def\ga{\gamma}
\def\o{\otimes}
\def\q{\quad}
\def\qq{\qquad}

\def\Res_#1{\hskip 1pt \oint_{#1=0}{d#1 \over 2\pi i}}
\def\VO{\Phi}
\def\VOt{\widetilde{\Phi}}
\def\hh{\widehat{\goth{h}}\hskip 2pt}
\def\K{F}

\def\head#1^#2{\buildrel {\scriptscriptstyle #2} \over #1}

\def\qbinom#1#2{{#1}\atopwithdelims[]{#2}}




\def\qq{\qquad}
\def\q{\quad}

\overfullrule=0pt

\rightline{RIMS-926}
\rightline{June 1993}
\vskip 7 truepc
\centerline{\bf
Correlation functions of the spin-1 analog of the XXZ model
}\par\vskip 2 truepc
\centerline{
Makoto  Idzumi$^{\dag}$
}\par\vskip 1 truepc
\centerline{\it
Research Institute for Mathematical Sciences,  Kyoto University,  Kyoto 606}
\footnote{}{\par
\noindent
$^{\dag}$On leave from Department of Applied Physics,
Faculty of Engineering,
The University of Tokyo,
Hongo, Bunkyo-ku, Tokyo 113
\par
}
\vskip 10 truepc

{\leftskip=1cm \rightskip=1cm
Exact integral representations of spin one-point functions
(ground state expectation values) are reported
for the spin-1 analog of the XXZ model in the region $-1<q<0$.
The method enables one to calculate arbitrary $n$-point functions
in principle.
We also report a construction of level 2 irreducible highest weight
representations of $\u$ in terms of boson and fermion operators,
and explicit forms of related vertex operators.
\par}

\vfill\eject

\def\refhat#1{$^{#1)}$}
\def\refno#1{#1}
\def\DFJMN{1}
\def\JMMN{2}
\def\IIJMNT{3}
\def\FrJ{4}
\def\CP{5}
\def\LepowskyP{6}
\def\Bernard{7}
\def\TsuchiyaK{8}
\def\FR{9}
\def\DJO{10}
\def\Ka2{11}
\def\DJMO{12}


\beginsection \S1. Introduction

Recent developments of a representation theoretical approach
to the XXZ quantum spin chain for $\Delta<-1$ ($-1<q<0$)
enable one not only to diagonalize the Hamiltonian
but also to calculate arbitrary spin correlation functions
exactly in the form of integral representations\refhat{\DFJMN,\JMMN}.
Higher spin analogs of the XXZ model could be attacked as well
and excitation sprectra were obtained\refhat{\IIJMNT}.
The method fully uses the representation theory of quantum affine algebras:
the XXZ model and its higher spin analogs has an exact symmetry $\u$
if the spin chain is infinite in both directions.
A space of states is identified with a tensor product of
an irreducible highest weight module over $\u$ of level $k$
(if the spin is $k/2$) and its {\it dual}.
Vertex operators play a key role in this identification.
The vacuum (the ground state), which should be unique if a boundary condition
is fixed, is identified with the unique one-dimensional submodule in the
space of states.
In terms of the vacuum and the vertex operators
we can write down an exact expression of an arbitrary
spin $n$-point correlation function (cf. eq.(4.2)),
where the correlation function means
the expectation value with respect to the vacuum.

In ref.\refno{\JMMN}, correlations of the XXZ model were calculated
through a concrete realization of the level 1 irreducible highest weight
modules over $\u$ in terms of boson\refhat{\FrJ}
(the procedure is called bosonization).
A success in obtaining arbitrary correlations in ref.\refno{\JMMN}
partly relies on the simplicity of the boson calculus.
If one tries to apply the same method to the higher spin problem,
he would realize that a way to the goal is not so easy to go ahead:
first he must know a concrete realization of
level $k$ irreducible highest weight representations
which is known very complicated for $k>1$,
he must find the explicit forms of vertex operators
related to the representations,
and finally using these vertex operators
he must perform calculations to get some simpler and more explicit expression
of correlation functions.

The present paper treats with the spin 1 analog of the XXZ model
in the region $-1<q<0$.
We construct the level 2 irreducible highest weight representations
explicitly in terms of boson, and Neveu-Schwarz and Ramond fermions,
and give explicit forms of vertex operators (\S3).
After giving exact expressions of arbitrary correlations
for spin $k/2$,
we specialize ourselves to calculations of spin one-point functions
for spin 1 (\S4).
Since we shall follow the previous works completely,
we omit to expose details of the approach
(cf. ref.\refno{\JMMN}; and refs.\refno{\DFJMN,\IIJMNT}).
The method is only sketched quickly at the beginning of \S4.
We prepare necessary notations in the next section and
report novel results in the subsequent sections.

\beginsection \S2. Notations

Notations for $\u$ follow the refs.\refno{\IIJMNT,\JMMN}.
We set $\K={\bf Q}(q)$ for a field
where a parameter $q$ is an indeterminate here
(but later in \S4 we shall
regard it as a complex number in a range $-1<q<0$).
Set $P^*=\Z h_0 + \Z h_1 + \Z d$,  $P=\Z \La_0 + \Z \La_1 + \Z \delta$,
$\hh = \K h_0 \oplus \K h_1 \oplus \K d$ (the Cartan sub-algebra) and
$\hh^{\ast} = \K \La_0 \oplus \K \La_1 \oplus \K \delta =
\K \alpha_0 \oplus \K \alpha_1 \oplus \K \La_0$
(The basis $\{ \La_0,\La_1,\delta \}$ of $\hh^*$ is dual to
$\{ h_0,h_1,d \}$ of $\hh$, and
$\alpha_1=2\La_1-2\La_0$, $\alpha_0=\delta-\alpha_1$).
Define a symmetric bilinear form on $\hh$ and the one induced on $\hh^*$ by
$$
(h_i,h_j)=\cases{2 & for $i=j$\cr -2 & for $i\not= j$\cr}, \q
(h_i,d)=\delta_{i,0}, \q (d,d)=0;
$$
$$
(\La_i,\La_j)={1\over 2}\delta_{i,1}\delta_{j,1}, \q
(\La_i,\delta)=1, \q (\delta,\delta)=0,
$$
$$
(\alpha_i,\alpha_j)=\cases{2 & for $i=j$\cr -2 & for $i\not= j$\cr}, \q
(\alpha_i,\La_0)=\delta_{i,0}, \q (\La_0,\La_0)=0.
$$
A quantum affine algebra $\u$ is an associative algebra with unit $1$
over a field $\K={\bf Q}(q)$
generated by $e_i$, $f_i$ ($i=0,1$), $q^h$ ($h\in P^*$)
with relations
$$
q^h q^{h'} = q^{h+h'}, \q
q^0 = 1, \q
q^h e_i q^{-h} = q^{\br{\alpha_i,h}} e_i, \q
q^h f_i q^{-h} = q^{-\br{\alpha_i,h}} f_i,
$$
$$
[e_i,f_j] = \delta_{ij} \ov{t_i-t_i^{-1}}{q-q^{-1}}\q (t_i=q^{h_i}),
$$
where $h,h'\in P^*$,  $i,j=0,1$, and
$$
\eqalignno{
e_i^3e_j-[3]e_i^2e_je_i+[3]e_ie_je_i^2-e_je_i^3 & =0 \q (i\not= j),\cr
f_i^3f_j-[3]f_i^2f_jf_i+[3]f_if_jf_i^2-f_jf_i^3 & =0 \q (i\not= j).\cr
}
$$
We have used and will use notations such as
$$
[x]={q^x-q^{-x} \over q-q^{-1}},\q [n]!=[n][n-1]\cdots[2][1],\q
{\qbinom{n}{k}}={[n]! \over [k]![n-k]!}.
$$
Let us define the coproduct and the antipode
on the generators by
$$
\Delta(e_i) = e_i\o 1 + t_i\o e_i, \q
\Delta(f_i) = f_i\o t_i^{-1} + 1\o f_i, \q
\Delta(q^h) = q^h\o q^h \q (h\in P^*);
$$
$$
a(e_i) = -t_i^{-1} e_i, \q
a(f_i) = -f_i t_i, \q
a(q^h) = q^{-h} \q (h\in P^*).
$$

The Drinfeld's realization of {\rm $\Up$},
which is a subalgebra of $\u$
generated by $\{ e_i, f_i, t_i | i=0, 1 \}$ only,
is defined as follows:
generators are
$x_m^{\pm} (m\in \Z)$, $a_m (m\in \Z_{\not= 0})$, $\ga$, and $K$,
and relations are
$$
\eqalignno{
& \ga \in \hbox{center}, \cr
& \big[ a_m, a_n \big] =
  \delta_{m+n,0} \ov{[2m]}{m} \ov{\ga^{m}-\ga^{-m}}{q-q^{-1}},
\q \big[ a_m, K \big] = 0, \cr
& K x_n^{\pm} K^{-1} = q^{\pm 2} x_n^{\pm},
\q \big[ a_m, x_n^{\pm} \big] =
  \pm \ov{[2m]}{m} \ga^{\mp{|m|\over 2}} x_{m+n}^{\pm}, \cr
& x_{m+1}^{\pm} x_{n}^{\pm} - q^{\pm 2} x_{n}^{\pm} x_{m+1}^{\pm} =
  q^{\pm 2} x_{m}^{\pm} x_{n+1}^{\pm} - x_{n+1}^{\pm} x_{m}^{\pm}, \cr
& \big[ x_{m}^{+}, x_{n}^{-} \big] =
  \ov{1}{q-q^{-1}} \big( \ga^{{1\over 2} (m-n)} \psi_{m+n} -
  \ga^{-{1\over 2} (m-n)} \varphi_{m+n} \big),
&(2.1)\cr}
$$
where
$$
\eqalignno{
\sum_{m=0}^{\infty} \psi_{m} z^{-m}
& = K \exp\Big( (q-q^{-1})\sum_{m=1}^{\infty} a_{m} z^{-m} \Big), \cr
\sum_{m=0}^{\infty} \varphi_{-m} z^{m}
& = K^{-1} \exp\Big( -(q-q^{-1})\sum_{m=1}^{\infty} a_{-m} z^{m} \Big), \cr
}
$$
and $\psi_{-m} = \varphi_{m} = 0$ for $m>0$.
The bracket~$[X,Y]$ means $XY-YX$.
We note thet
$a_m$ ($m\in\Z_{\not= 0}$) and the center generated by $\gamma$
form a Heisenberg subalgebra with respect to the bracket~$[\q,\q]$.
We therefore regard this $a_m$ as boson.
The Chevalley generators $\{e_i,f_i,t_i\}$ of $\Up$
are given by the identification
$$
\eqalignno{
&
t_0=\ga K^{-1}, \q
t_1= K, \q
e_1= x_0^+, \q
f_1= x_0^-, \q
e_0t_1= x_1^-, \q
t_1^{-1}f_0=x_{-1}^+.
&(2.2)\cr
}
$$
The coproduct of the Drinfeld generators is known partially\refhat{\CP}:
for $k\geq0$ and $l>0$ we have
$$
\eqalignno{
& \Delta(x_k^{+})=x_k^{+}\o\ga^k+\ga^{2k}K\o x_k^{+}+
  \sum_{i=0}^{k-1}\ga^{(k+3i)/2}\psi_{k-i}\o\ga^{k-i}x_i^{+}
  \qq \bmod N_{-}\o N_{+}^2,
\cr
& \Delta(x_{-l}^{+})=x_{-l}^{+}\o\ga^{-l}+K^{-1}\o x_{-l}^{+}+
  \sum_{i=1}^{l-1}\ga^{(l-i)/2}\varphi_{-l+i}\o\ga^{-l+i}x_{-i}^{+}
  \qq \bmod N_{-}\o N_{+}^2,
\cr
& \Delta(a_l)=a_l\o\ga^{l/2}+\ga^{3l/2}\o a_l
  \qq \bmod N_{-}\o N_{+},
\cr
& \Delta(a_{-l})=a_{-l}\o\ga^{-3l/2}+\ga^{-l/2}\o a_{-l}
  \qq \bmod N_{-}\o N_{+}.
\cr
& &(2.3)\cr
}
$$
Here $N_{\pm}$ and $N_{\pm}^2$ are left
$\K[\ga^\pm,\psi_r,\varphi_s|r,-s\in{\bf Z}_{\geq0}]$-modules
generated by $\{x_m^{\pm}|m\in {\bf Z} \}$ and
$\{x_m^{\pm}x_n^{\pm}|m,n\in {\bf Z}\}$ respectively.
It gives sufficient information for our calculation.

Making use of the coproduct and the antipode,
we can define canonically the tensor product and the dual
representations of $\u$:
(i) given representations $(\pi_V,V)$, $(\pi_W,W)$,
the tensor product representation $(\pi_{V\o W},V\o W)$ is defined to be
$\pi_{V\o W}=(\pi_V \o \pi_W)\circ \Delta$;
(ii) given a representation $(\pi,V)$,
the dual representation $(\pi^{\ast a^{\pm 1}},V^{\ast})$ is defined to be
$\pi^{\ast a^{\pm 1}}={}^{t}\pi\circ a^{\pm 1}$.
We also denote them $V^{\ast a^{\pm 1}}$ as modules over $\u$
with the action given by $\pi^{\ast a^{\pm 1}}$.
Note that they are left $\u$-modules.

Set $V^{(k)}=\K u_0 \oplus \K u_1 \oplus \cdots \oplus \K u_k$ and
$V^{(k)}_z=V^{(k)} \o \K[z,z^{-1}]$.
Then, equations
$$
\pi(t_1) u_{j} = q^{k-2j} u_{j}, \q
\pi(e_1) u_{j} = [j] u_{j-1}, \q
\pi(f_1) u_{j} = [k-j] u_{j+1},
$$
$$
\pi(t_0)=\pi(t_1)^{-1},\q \pi(e_0)=\pi(f_1),\q \pi(f_0)=\pi(e_1)
$$
or
$$
\pi(\ga) u_{j} = u_{j}, \q
\pi(K) u_{j}   = q^{k-2j} u_{j}, \q
\pi(a_m) u_{j} = a_m^j u_{j},
$$
$$
\pi(x_m^+) u_{j} = q^{m(k-2j)} [j] u_{j-1}, \q
\pi(x_m^-) u_{j} = q^{m(k-2j)} [k-j] u_{j+1},
$$
where
$$
a_m^j=-\ov{[2m][jm]}{m[m]} q^{m(k-j+1)} + \ov{[km]}{m},
$$
define
a $(k+1)$-dimensional (spin-$k/2$) representation
$(\pi,V^{(k)})$ of $\Up$.
Equations
$$
\pi_z(x)=\pi(x)\o \hbox{id} \q \hbox{for} \q x=e_1, f_1, t_1, t_0,
$$
$$
\pi_z(e_0)=\pi(f_1)\o z, \q \pi_z(f_0)=\pi(e_1)\o z^{-1},
$$
$$
\pi_z(q^d) (u_j \o z^n) = q^n  u_j \o z^n
$$
define a representation $(\pi_z,V^{(k)}_z)$ of $\u$.
In terms of the Drinfeld generators,
the action of $\Up$ on $V^{(k)}_z$ is written as
$$
\pi_z(\ga) = \pi(\ga)\o \id, \q
\pi_z(K)   = \pi(K)\o \id, \q
\pi_z(a_m) = \pi(a_m)\o z^m,
$$
$$
\pi_z(x_m^+) = \pi(x_m^+)\o z^m, \q
\pi_z(x_m^-) = \pi(x_m^-)\o z^m.
$$

\beginsection \S3. Level 2 irreducible highest weight representations of $\u$
\par
{\bf   and vertex operators}\par\bigskip

\noindent
Set $P_+ = \Z_{\ge 0} \La_0 \o \Z_{\ge 0} \La_1$.
For $\la\in P_+$, a $\u$-module $V(\la)$ is called
an irreducible highest weight module with highest weight $\la$
if the following conditions are satisfied:
there is a vector $\ket{\la}\in V(\la)$, called the highest weight vector,
and
$q^h \ket{\la} = q^{\br{\la,h}} \ket{\la} \q (h\in P^*)$,
$e_i \ket{\la} = 0$, $f_i^{\br{\la,h_i}+1} \ket{\la} = 0 \q (i=0,1)$,
and $V(\la) = \u \ket{\la}$.
We say that $V(\la)$ has level~$k$ if $\br{\la,c}=k (\in\Z_{\ge 0})$.
The $V(\la)$ has a weight-space decomposition
$V(\la)=\bigoplus_{\xi\in P} V(\la)_\xi$.
We sometimes write a weight of a weight vector $v$ as $\wt(v)$.

To construct the level 2 irreducible highest weight modules,
we may need to introduce some fermions,
since we know that we really have to introduce fermions
to construct the corresponding level 2 modules over
an affine Lie algebra $\slth$
(see, for instance, ref.\refno{\LepowskyP}).
Incidentally,
there exists a work by Bernard\refhat{\Bernard} on
level~$1$ representations of $U_q(B^{(1)}_r)$.
It is helpful to us
since $o_3 \simeq sl_2$ and $B^{(1)}_r \simeq \widehat{o}_{2r+1}$.
So, let us construct our modules refering to these references as a guide.


Let $P={\bf Z}{\alpha\over2}$, $Q={\bf Z}\alpha$
be the weight/root lattice of $\slt$, and let
$\K[P]$, $\K[Q]$ be their group algebras (
Do not confuse this $P$ with $P=\Z \La_0+\Z \La_1+\Z \delta$,
and this $Q$ with
the field of rational numbers).
The basis elements of $\K[P]$ are written
multiplicatively as $\e^{n\alpha}$ $(n\in {1\over2}{\bf Z})$.
Let ${\cal F}^{a}=\K[a_{-1},a_{-2},\cdots]$ be the boson Fock space.

In contrast to the level 1 case,
we need furthermore two species of fermions
to construct the {\it boson vacuum space} in $V(\la)$.
Define
$$
\phi^{NS}(z)=\sum_{n\in\Z+{1\over 2}}\phi_n z^{-n}, \q
\phi^{R}(z)=\sum_{n\in\Z}\phi_n z^{-n}
$$
by the anticommutation relations
$$
\big[ \phi_m,\phi_n \big]_+ =
  \delta_{m+n,0} \ov{q^{2m}+q^{-2m}}{q+q^{-1}}.
$$
We call $\{\phi_n | n\in\Z+{1\over 2}\}$  NS fermions (after Neveu-Schwarz),
and $\{\phi_n | n\in\Z\}$  R fermions (after Ramond).
Let ${\cal F}^{\phi^{NS}}=\K[\phi_{-1/2},\phi_{-3/2},\cdots]$ and
${\cal F}^{\phi^{R}} = {\cal F}^{\psi} \o \C^2
= \K[\psi_{-1},\psi_{-2},\cdots] \o \C^2$
be the fermion Fock spaces.
\def\MatZ{\left(\matrix{1&0\cr 0&-1\cr}\right)}
\def\MatI{\left(\matrix{0&1\cr 1&0\cr}\right)}
Here we have set for R fermion $\phi_n = \psi_n \o \MatZ$ for $n\not= 0$
and $\phi_0 = \psi_0 \o \MatI$,
where $\psi_n$ ($n\not= 0$) satisfy the same anti-commutation relations
as $\phi_n$
and $\psi_0 = 1/(q+q^{-1})^{1/2}$ is an ordinary number
(i.e., $\psi_0$ commutes with all $\psi_n$, $n \not= 0$),
because
the total Fock space of R fermion $\{\phi_n\}$ decomposes
into two mutually isomorphic sectors.

Now we can define the action of $\Up$
on spaces
$W={\cal F}^{a} \o {\cal F}^{\phi^{i}} \o \K[P]$ ($i=NS,R$).
The latter two components form the boson vacuum space.
The bosons $\{a_m\}$ act on the first component of $W$ as
$$
\eqalignno{
a_m(f\o g\o\e^{\beta})&=a_m f \o g\o \e^{\beta} \qq \hbox{for} \q m<0, \cr
a_m(f\o g\o\e^{\beta})&=[a_m, f] \o g\o \e^{\beta} \qq \hbox{for} \q m>0. \cr
}
$$
Define operators $\e^\beta$ and $\partial_\alpha$, $\beta \in \C\alpha$,
by
$$
\eqalignno{
\e^{\beta_1}(f\o g\o\e^{\beta_2})&=f\o g\o\e^{\beta_1+\beta_2}, \cr
\partial_\alpha(f\o g\o\e^{\beta})&=(\alpha,\beta)f\o g\o\e^{\beta}. \cr
}
$$
The fermions  $\{\phi_m\}$ act on $W$ as
$$
\phi_m(f\o g\o\e^{\beta})=f \o (\phi_m g) \o \e^{\beta}
$$
with $\phi_m(f\o 1\o\e^{\beta})=0$ for $m>0$.
We let also $K=1\o 1\o q^{\partial_\alpha}$ and
$$
\gamma=q^2 \o 1 \o 1.
$$
(Note that the element $c\in P^{\ast}$ act as scalar $2$ on the level~$2$
module and that the $\gamma$ is regarded as $q^c$.)
Then,
from the Drinferd's relations (2.1),
forms of generating functions (often called the currents)
$x^{\pm}(z)=\sum_{n\in{\bf Z}}x_n^{\pm}z^{-n}$
acting on $W$ are determined as
$$
\eqalignno{
x^+(z)&=
\eta
\exp\bigl(\sum_{m=1}^\infty{a_{-m}\over[2m]}q^{-m}z^m\bigr)
\exp\bigl(-\sum_{m=1}^\infty{a_{m}\over[2m]}q^{-m}z^{-m}\bigr)
\phi(z)
\e^\alpha z^{{1\over 2}+{1\over 2}\partial_\alpha},
\cr
x^{-}(z)&=
\eta
\exp\bigl(-\sum_{m=1}^\infty{a_{-m}\over[2m]}q^{m}z^m\bigr)
\exp\bigl(\sum_{m=1}^\infty{a_{m}\over[2m]}q^{m}z^{-m}\bigr)
\phi(z)
\e^{-\alpha} z^{{1\over 2}-{1\over 2}\partial_\alpha},
\cr
}
$$
where $\eta=(q+q^{-1})^{1/2}$
(Our definition of currents is differnt from the one in ref.\refno{\JMMN}).
With these actions we can obtain the irreducible modules.
Let ${\cal F}_{even/odd}$ be subspaces of the Fock space ${\cal F}$
consisting of even/odd particle states.
Then the followings are $\Up$-modules:
$$
\eqalignno{
V(2\La_0)
&\simeq
 \bigl( {\cal F}^a \o {\cal F}^{\phi^{NS}}_{even} \o
 \K[2Q]\bigr)
 \oplus
 \bigl( {\cal F}^a \o {\cal F}^{\phi^{NS}}_{odd} \o
 \e^\alpha \K[2Q]\bigr), \cr
V(2\La_1)
&\simeq
 \bigl( {\cal F}^a \o {\cal F}^{\phi^{NS}}_{even} \o
 \e^\alpha \K[2Q]\bigr)
 \oplus
 \bigl( {\cal F}^a \o {\cal F}^{\phi^{NS}}_{odd} \o
 \K[2Q]\bigr), \cr
}
$$
$$
\eqalignno{
V(& \La_0+\La_1) \cr
&\simeq
 \bigl( {\cal F}^a \o
 {\cal F}^{\psi}_{even} \o {1\choose 1} \o
 \e^{\alpha\over 2} \K[Q]\bigr)
 \oplus
 \bigl( {\cal F}^a \o
 {\cal F}^{\psi}_{odd} \o {1\choose -1} \o
 \e^{\alpha\over 2} \K[Q]\bigr) \cr
&{\buildrel \hbox{or} \over \simeq }
 \bigl( {\cal F}^a \o
 {\cal F}^{\psi}_{odd} \o {1\choose 1} \o
 \e^{\alpha\over 2} \K[Q]\bigr)
 \oplus
 \bigl( {\cal F}^a \o
 {\cal F}^{\psi}_{even} \o {1\choose -1} \o
 \e^{\alpha\over 2} \K[Q]\bigr). \cr
}
$$
The highest weight vectors are $1\o 1\o 1$ for $V(2\La_0)$,
$1\o 1\o \e^\alpha$ for $V(2\La_1)$,
and
$1\o \bigl(1\o{1\choose 1}\bigr)\o \e^{\alpha\over 2}$ or
$1\o \bigl(1\o{1\choose -1}\bigr)\o \e^{\alpha\over 2}$
for $V(\La_0+\La_1)$ according to the two realizations,
respectively.

We can define the grading operator $d$ of $\u$  by
$$
\eqalignno{
d (& a_{-m_1}^{n_1}\cdots a_{-m_r}^{n_r}\o\phi_{-k_1}\cdots\phi_{-k_s}v
\o\e^\beta)
\cr
&=\bigl(-\sum_{j=1}^rm_jn_j-\sum_{j=1}^sk_j
-{(\beta,\beta)\over4}+{(\la,\la)\over4}\bigr)
(a_{-m_1}^{n_1}\cdots a_{-m_r}^{n_r}\o\phi_{-k_1}\cdots\phi_{-k_s}v
\o\e^\beta)
\cr
&=\bigl(-\sum_{m=1}^{\infty}mN_m^a-\sum_{k>0}kN_k^{\phi}
-{1\over8}\partial_\alpha^2+{(\la,\la)\over4}\bigr)
(a_{-m_1}^{n_1}\cdots a_{-m_r}^{n_r}\o\phi_{-k_1}\cdots\phi_{-k_s}v
\o\e^\beta)
\cr
}
$$
on $V(\la)$,
where
the boson and the fermion number operators are defined by (for $m>0$)
$$
\eqalignno{
N_m^a&={1\over [a_m,a_{-m}]}a_{-m}a_m={m\over [2m]^2}a_{-m}a_m,
\cr
N_m^{\phi}&=
{1\over [\phi_m,\phi_{-m}]_+}\phi_{-m}\phi_m=
{q+q^{-1}\over q^{2m}+q^{-2m}}\phi_{-m}\phi_m.
\cr
}
$$
With this definition of the action of $d$, the $V(\la)$ become
irreducible highest weight $\u$-modules.

We checked the validity of the expressions for currents $x^{\pm}(z)$
for `higher' $26$ weight vectors
(i.e., vectors $f_{i_r} \cdots f_{i_1} \ket{\la}$ with $r$ small,
$\ket{\la}$ the highest weight vector)
of $V(2\La_i)$, $i=0,1$,
and $35$ weight vectors of $V(\La_0 + \La_1)$;
we checked the defining relations for the Chevalley generators
on these vectors.

At this point,
let us define the normal-ordered product of two fermion fields.
We want to define it by
$$
\eqalignno{
{\cal N}[\phi_m \phi_n] &= \phi_m \phi_n
    \qquad\hbox{for $m,n\ge 0$ or $m,n\le 0$}; \cr
{\cal N}[\phi_m \phi_{-n}] &= -\phi_{-n} \phi_m
    \qquad\hbox{for $m,n> 0$}; \cr
{\cal N}[\phi_{-m} \phi_{n}] &= \phi_{-m} \phi_n
    \qquad\hbox{for $m,n> 0$}. \cr
}
$$
For the fermion fields $\phi(z)$, we set
${\cal N}[\phi(z) \phi(w)] =
 \sum\sum {\cal N}[\phi_m \phi_n] z^{-m} w^{-n}$.
This definition yields
${\cal N}[\phi(z) \phi(z)] = 0$ for NS fermion;
but for R fermion ${\cal N}[\phi(z) \phi(z)] = (\phi_0)^2$.
So we re-define the normal order product of the fermion fields as
$$
:\phi(z)\phi(w): = {\cal N}[\phi(z) \phi(w)] - \varepsilon (\phi_0)^2,
$$
$$
\varepsilon=\cases{0& for NS\cr
                   1& for R.\cr}
$$
The propagator has the following form:
$$
\eqalignno{
\br{\phi(w_1)\phi(w_2)}
&{\buildrel \hbox{def} \over =} \phi(w_1)\phi(w_2) - : \phi(w_1)\phi(w_2) : \cr
&=\ov{\sigma(w_2/w_1)(1-w_2/w_1)}{(1-q^2w_2/w_1)(1-q^{-2}w_2/w_1)}, \cr
}
$$
$$\sigma(w_2/w_1)=\cases{(w_2/w_1)^{1/2}& for NS\cr
                         {1\over q+q^{-1}}(1+w_2/w_1)& for R.\cr}
$$
It is defined in a region $|q^{2}w_2/w_1|,|q^{-2}w_2/w_1|<1$.
We note a relation
$$
\eqalignno{
\ov{\br{\phi(w_1)\phi(w_2)}}{w_1(1-w_2/w_1)}
&-
\ov{\br{\phi(w_2)\phi(w_1)}}{w_2(1-w_1/w_2)}
\cr
= &
\ov{\delta(q^2w_2/w_1)}{q^{-1}(q^2-q^{-2})w_1} -
\ov{\delta(q^{-2}w_2/w_1)}{q(q^2-q^{-2})w_1}.
\cr
}
$$

Now let us find expressions for vertex operators acting on them.
They are the $q$-deformation of the vertex operators of
Tsuchiya-Kanie type\refhat{\TsuchiyaK},
and are first introduced by Frenkel-Reshetikhin\refhat{\FR}.
We write level $k$ highest weights as $\la_m=(k-m)\La_0+m\La_1$,
$m=0,1,\ldots,k$.
For two level $k$ highest weight modules $V(\la_{m})$, $V(\la_{k-m})$ and
the $(k+1)$-dimensional representation $V$,
consider a map
$$
\VO_{\la_{m}}^{\la_{k-m}V}(z):  V(\la_{m})\longrightarrow V(\la_{k-m})\o V_z
$$
acting as an intertwiner
$$
\eqalignno{
&\Delta(x)\circ \VO(z) = \VO(z) \circ x  \q \hbox{for } x\in\u.
&(3.1)
\cr}
$$
This is a vertex operator.
Precisely speaking,  it is defined  as an intertwiner
$\VO(z):  V(\la_{m})  \longrightarrow   V(\la_{k-m})  \widehat{\otimes} V_z$
where
$M \widehat{\otimes}  N =  \bigoplus_\xi  \prod_\nu  M_\nu \o N_{\xi-\nu}$.
It exists uniquely up to normalization for each $m = 0,1, \ldots, k$
(The existence is proven in ref.\refno{\DJO} for more general setting).
We normalize it as
$$
\eqalignno{
&\VOt_{\la_{m}}^{\la_{k-m} V}(z)\bigl(\ket{\la_{m}}\bigr) =
\ket{\la_{k-m}}\o u_{k-m} + \cdots.
&(3.2)
\cr}
$$
Here and in the following,
we will use the notation $\VOt$ for normalized vertex operators
(such as (3.2)),
and $\VO$ for the others (unnormalized ones).
Note that for level~$k$
we have $(k+1)$~vertex operators (for $m=0,1,\ldots,k$)
of this type.

Let us determine the explicit form of $\VO(z)$ which satisfies the
intertwining relations (3.1).
We set
$$
\VO(z)=\sum_{j=0}^{k} \VO_j(z)\o u_j
$$
and call $\VO_j(z)$ the $j$-th component of $\VO(z)$.

{}From the intertwining relation with $f_1$,
we get relations among $\VO_j$ such as
$$
\eqalignno{
\VO_0(z)f_1 &= q^{-k}f_1\VO_0(z), \cr
\VO_j(z)f_1 &= q^{2j-k} f_1\VO_j(z) + [k-j+1]\VO_{j-1}(z),\q j=1,\ldots,k.\cr
& &(3.3)\cr
}
$$
The second equations yield
$$
\eqalignno{
\VO_{j-1}(z)
&= {1\over [k-j+1]}\bigl(\VO_j(z)f_1-q^{2j-k}f_1\VO_j(z)\bigr) \cr
&= {1\over [k-j+1]}\Res_{{w}}
   {1\over w}\bigl(\VO_j(z)x^-(w)-q^{2j-k}x^-(w)\VO_j(z)\bigr) \cr
}
$$
for $j=1,\ldots,k$,
since $f_1 = x_0^-=\Res_{{w}} (x^-(w)/w)$.
The contour around $w=0$ is chosen
such that
the integrand is convergent wherein.

In order to find an expression of $\VO_k(z)$,
we  use (3.1) for $x=a_m$, $x_n^+$ and $K$
(the coproduct of these generators are given in
eq.(2.3)).
The result is
$$
\eqalignno{
&a_m\VO_k(z)-\VO_k(z)a_m
=q^{({3k\over2}+2)m} {[km] \over m} z^m \VO_k(z) \q \hbox{for $m>0$},
&(3.4a)\cr
&a_{-m}\VO_k(z)-\VO_k(z)a_{-m}
=q^{-({k\over2}+2)m} {[km] \over m} z^{-m} \VO_k(z) \q \hbox{for $m>0$},
&(3.4b)\cr
&\VO_k(z)x^+(w)-x^+(w)\VO_k(z)=0,
&(3.4c)\cr
&K\VO_k(z) K^{-1}=q^k\VO_k(z).
&(3.4d)\cr
}
$$
There is another equation
$$
\eqalignno{
&
\sum_{l=0}^{k+1} (-1)^l {\qbinom{k+1}{l}} f_1^l \VO_k(z) f_1^{k+1-l}
= 0
\cr
}
$$
obtained from (3.3),
which, however, we shall not use.

{}From these relations, we propose the following explicit forms
of vertex operators.

\proclaim Proposition 1.
An operator acting on the space ${\cal F}^a\o {\cal F}^\phi\o\K[P]$
with components
$$
\VO_2(z)=
\exp\bigl(\sum_{m=1}^\infty{a_{-m}\over[2m]}q^{5m}z^m\bigr)
\exp\bigl(-\sum_{m=1}^\infty{a_{m}\over[2m]}q^{-3m}z^{-m}\bigr)
\e^{\alpha} (-q^4z)^{{1\over 2}\partial_\alpha},
$$
$$
\eqalignno{
\VO_1(z)&=
\Res_{{w_1}}\ov{1}{w_1}\bigl(\VO_2(z)x^-(w_1)-q^2x^-(w_1)\VO_2(z)\bigr) \cr
&=
-\ov{1-q^4}{q^4z}\Res_{{w_1}}\ov{1}{w_1(1-q^{-2}w_1/z)(1-q^6z/w_1)}
: \VO_2(z)x^-(w_1) : , \cr
}
$$
$$
\eqalignno{
&\VO_0(z)=
\ov{1}{[2]} \Res_{{w_2}}\ov{1}{w_2}
\bigl(\VO_1(z)x^-(w_2)-x^-(w_2)\VO_1(z)\bigr)
\cr
&\qquad
=
\Res_{{w_2}}\Res_{{w_1}} \Bigl\{
\cr
&\quad
\ov{1}{[2]} \Big( \ov{1-q^4}{q^4z} \Big)^2
\ov{(1-q^4z/w_1)}{(1-q^{-2}w_1/z)(1-q^6z/w_1)w_2(1-q^{-2}w_2/z)(1-q^6z/w_2)}
\cr
&\qquad\quad\times
: \VO_2(z)x^-(w_1)x^-(w_2) : \cr
&\quad\quad +
\ov{1}{[2]} \ov{1-q^4}{(q^4z)^2}
\ov{(1-q^2w_2/w_1) \br{\phi(w_1)\phi(w_2)}}
{(1-q^{-2}w_1/z)(1-q^6z/w_1)w_2(1-q^{-2}w_2/z)}
\cr
&\qquad\quad\times
: \VO_2(z)\widehat{x^-}(w_1)\widehat{x^-}(w_2) : \cr
&\quad\quad +
\ov{1}{[2]} \ov{1-q^4}{q^4z}
\ov{(1-q^2w_1/w_2)\br{\phi(w_2)\phi(w_1)}}
{w_1(1-q^{-2}w_1/z)(1-q^6z/w_1)w_2(1-q^6z/w_2)} \cr
&\qquad\quad\times
: \VO_2(z)\widehat{x^-}(w_1)\widehat{x^-}(w_2) :
\Bigr\} \cr
}
$$
$$
\eqalignno{
=
& \Res_{{w_2}}\Res_{{w_1}} \Biggl\{
\ov{1}{[2]} \ov{1-q^4}{q^8z^3} \cdot
\ov{1}{{w_1\over z}{w_2\over z}(1-q^{-2}{w_1\over z})(1-q^{-2}{w_2\over z})
(1-q^6{z\over w_1})(1-q^6{z\over w_2})}
\cr
&\quad\times \biggl[
(1-q^4){w_1\over z}(1-q^4{z\over w_1}) : \VO_2(z)x^-(w_1)x^-(w_2) : \cr
&\quad\quad + \Bigl[
{w_1\over z}(1-q^6{z\over w_2})(1-q^2{w_2\over w_1})\br{\phi(w_1)\phi(w_2)}
\cr
&\quad\quad\quad +
q^4(1-q^{-2}{w_2\over z})(1-q^2{w_1\over w_2})\br{\phi(w_2)\phi(w_1)}
\Bigr]
: \VO_2(z)\widehat{x^-}(w_1)\widehat{x^-}(w_2) : \biggr]
\Biggr\} \cr
}
$$
satisfies the relations (3.1),
where $:$ $:$ denotes the normal ordering with respect to
boson, fermion and lattice operators
\footnote{$^{\dag}$}{
{\rm
For bosons
$$
\eqalignno{
:a_{m} a_{-n}:
&=a_{-n} a_{m} + \delta_{mn}c_m \q (m,n>0); \cr
:a_{-m} a_{n}:
&=a_{-m} a_{n}; \q (m,n>0;\hbox{already normal-ordered}); \cr
:a_{m} a_{n}:
&=a_{m} a_{n} =a_{n} a_{m} \q (m,n>0 \hbox{ or } m,n<0). \cr
}
$$
For lattice
$$
\eqalignno{
:z^{\partial_\alpha} \e^{\beta}:
&= z^{(\alpha,\beta)} \e^{\beta}z^{\partial_\alpha} \q (\beta=x \alpha); \cr
:\e^{\beta} z^{\partial_\alpha} :
&= \e^{\beta} z^{\partial_\alpha} \q (\hbox{already normal-ordered}). \cr
}
$$
}
},
and the hat "$\; \widehat{} \;$'' denotes fermion contraction
$$
\widehat{x^{-}}(w)=
\eta
\exp\bigl(-\sum_{m=1}^\infty{a_{-m}\over[2m]}q^{m}w^m\bigr)
\exp\bigl(\sum_{m=1}^\infty{a_{m}\over[2m]}q^{m}w^{-m}\bigr)
\e^{-\alpha} w^{{1\over 2}-{1\over 2}\partial_\alpha}.
$$
The normalized vertex operators (3.2) are given by
$$
\eqalignno{
\VOt_{\la_{m}}^{\la_{2-m}V}(z) &= (-q^4z)^{m/2}\VO(z), \q m=0,2; \cr
\VOt_{\la_{1}}^{\la_{1}V}(z) &= \epsilon (-q^4z)^{1/2}\VO(z)
&(3.5)\cr
}
$$
where the vertex operator $\VOt_{\la_{1}}^{\la_{1}V}(z)$
acts on
the realization of $V(\la_1)  =  V(\La_0+\La_1)$
with highest weight vector $p_\epsilon \o \e^{{\alpha\over 2}}$,
where
$p_\epsilon = {1\choose \epsilon}$, $\epsilon = \pm 1$.

\Proof
First of all, we note that we have made everything in the normal order
in the above equations (cf. Appendix C).
The proposed vertex operators satisfy all the equations (3.4a--d).
In order to confirm that $\VO$ thus obtained are correct ones,
we must check the intertwining relations (3.1)
at least for the Chevalley generators (2.2).
We checked the intertwining relations partially, but not all.
So we compared matrix elements of $\VO_2$ to the correct ones,
which are known for `higher' weight vectors
(i.e., vectors $f_{i_r} \cdots f_{i_1} \ket{\la}$ with $r$ small,
$\ket{\la}$ the highest weight vector)
and obtained using the global crystal base of Kashiwara\refhat{\Ka2}.
Actually we checked
`higher'~$26 \times 17$ matrix elements
(including many $0$ elements) for $\VOt^{2\La_{1-i} V}_{2\La_i}$ ($i=0,1$)
and `higher'~$35 \times 15$ matrix elements
for $\VOt^{\La_0+\La_1 V}_{\La_0+\La_1}$.
We found that they are correct.
Thus, though the proposition  is not completely proven
we think that it must be  correct certainly.
\hfill\qed

The {\it inverse} vertex operator
$$
\VOt_{\la_{k-m}V}^{\la_{m}}(z): V(\la_{k-m})\o V_z \longrightarrow V(\la_{m})
$$
which is an intertwiner and normalized as
$$
\eqalignno{
&\VOt_{\la_{k-m}V}^{\la_{m}}(z)\bigl(\ket{\la_{k-m}}\o u_{k-m}\bigr) =
\ket{\la_{m}} + \cdots
&(3.6)
\cr}
$$
can be obtained from the previous one as follows.
First let us identify a vertex operator
$V(\mu)\o V_z \rightarrow V(\la)$ with
$V(\mu) \rightarrow V(\la)\o V_z^{\ast a}$
$$
\VO_{\mu V,j}^{\la}(z)=\VOt_{\mu,j}^{\la V^{\ast a}}(z), \q j=0,1,\ldots,k,
$$
where
we write
$$
\eqalignno{
\VOt_{\mu}^{\la V^{\ast a}}(z)
&=\sum_{j=0}^k \VOt_{\mu,j}^{\la V^{\ast a}}(z) \o u_j^{\ast}, \cr
\VO_{\mu V,j}^{\la}(z)(v)
&=\VO_{\mu V}^{\la}(z)(v \o u_j). \cr
}
$$
Second,  we must note an isomorphism between left $\u$-modules
(the {\it charge conjugation})
$$
\matrix{
V_{zq^{-2}} & \goto{\sim} & V_{z}^{\ast a} \cr
u_j         & \longmapsto & c_j u_{k-j}^{\ast}, \cr
}
$$
$j=0,1,\ldots,k$,
where $\hbox{dim}V=k+1$ and
$$
c_j=(-1)^jq^{j^2+(1-k)j}\ov{1}{\left[\matrix{k\cr j\cr}\right]}
$$
(for level 2, $c_0=1$, $c_1=-1/[2]$, $c_2=q^2$).
This gives
$$
\VOt_{\mu,j}^{\la V^{\ast a}}(z)=c_{k-j}\VOt_{\mu,k-j}^{\la V}(zq^{-2}).
$$
These two equations determine the (unnormalized) vertex operators.
The normalized vertex operators are
$$
\eqalignno{&
\VOt_{\la_{k-m} V, j}^{\la_{m}}(z)=
{c_{k-j}\over c_{m}} \VOt_{\la_{k-m},k-j}^{\la_{m} V}(zq^{-2}),
\q m=0,1,\ldots,k.
&(3.7)\cr
}
$$

We said that the $\VOt_{\la_{k-m} V}^{\la_{m}}(z)$
is the inverse of $\VOt^{\la_{k-m} V}_{\la_{m}}(z)$,
since
the following relations hold:
$$
\eqalignno{
\VOt_{\la_{k-m} V}^{\la_{m}}(z)\circ\VOt_{\la_{m}}^{\la_{k-m} V}(z)
&= g_{\la_{m}} \times \id_{V(\la_{m})}, \cr
\VOt_{\la_{m}}^{\la_{k-m} V}(z)\circ\VOt_{\la_{k-m} V}^{\la_{m}}(z)
&= g_{\la_{m}} \times \id_{V(\la_{k-m})\o V}, \cr
}
$$
where
$$
g_{\la_m}=q^{(k-m)m} {\qbinom{k}{m}}
{(q^{2(k+1)};q^4)_\infty \over (q^2;q^4)_\infty},
$$
$$
(x;p)_m = \prod_{j=1}^m (1-xp^{j-1}), \q
(x;p)_{\infty} = \prod_{j=1}^{\infty} (1-xp^{j-1}).
$$
For level $2$ it is
$$
g_{\la_{m}}=
\cases{\ov{1}{1-q^2} & for $m=0,2$ \cr
     \ov{1+q^2}{1-q^2} & for $m=1$. \cr}
$$
For general~$k$, they are proved in ref.\refno{\IIJMNT}
using a solution to the $q$-KZ equation.
In the present case ($k=2$),
since we have explicit expressions for the vertex operators,
they can be proved by direct calculation
(Since the intertwining relations hold,
it is sufficient to prove the relations
on the highest weight vector).

\beginsection \S4. Integral representations of spin correlation functions

First
we describe a general formulation for general $k$ quickly
(cf. refs.\refno{\DFJMN,\IIJMNT}).
The spin~$k/2$ Hamiltonian is defined as (cf. Appendix A)
$$
H=\sum_{l\in\Z} h_{l+1,l},
$$
$$
h_{l+1,l}=
\cdots \o \head{1}^{l+2}\o \head{h}^{l+1\q l} \o \head{1}^{l-1}\o \cdots,
$$
$$
\eqalignno{&
h=(-1)^k(q^k-q^{-k}) \biggl[{d\over dz}\Rc(z,1)\biggr]_{z=1}
&(4.1)\cr
}
$$
where
$$
{\check R}(z_1/z_2): V_{z_1}^{(k)}\o V_{z_2}^{(k)} \longrightarrow
V_{z_2}^{(k)}\o V_{z_1}^{(k)}
$$
is
the $R$-matrix which intertwins two $\u$-modules.
The Hamiltonian formally acts on an
infinite tensor product
$$
V^{\o\infty}=\cdots \o \head{V}^{2} \o \head{V}^{1} \o \head{V}^{0} \o
\head{V}^{-1} \o \cdots
$$
of the $(k+1)$-dimensional space $V=V^{(k)}$,
and has a symmetry
$$
[H_{XXZ},\Up]=0.
$$
In the following we assume $-1<q<0$,
and then $\Delta <-1$ ($\Delta=(q+q^{-1})/2$).
The basic idea in refs.\refno{\DFJMN,\IIJMNT} is
an identification of the space of states with some fixed boundary condition,
which is a subspace of $V^{\o\infty}$,
with a level-zero $\u$-module
$V(\la)\o V(\mu)^{\ast a}$.
This space is embedded into the infinite product space via vertex operators
$$
V(\la)\o V(\mu)^{\ast a}\hookrightarrow V^{\o\infty}.
$$
Specializing the level~$k$ highest weight $\la$
and the level~$-k$ lowest weight $\mu$
corresponds to fixing a boundary condition.
This boundary condition depends on the parameter $q$.
In the limit $q\to 0$ (which is an Ising limit; cf. Appendix A)
it is explicitly described as follows:
any vector in  $V(\la_m)\o V(\la_{m'})^{\ast a}$
tends to a pure tensor of the form
$\cdots\o u_{p(2)}\o u_{p(1)}\o u_{p(0)}\o u_{p(-1)}\o\cdots$
(when embedded into $V^{\o\infty}$)
with a boundary condition
$$
\eqalignno{
p(l)
&=\left\{\matrix{
m&\hbox{if }l\equiv 0\bmod 2 \cr
k-m&\hbox{if }l\equiv 1\bmod 2 \cr
}\right\}\quad
\hbox{for } l\gg 1, \cr
&=\left\{\matrix{
m'&\hbox{if }l\equiv 0\bmod 2 \cr
k-m'&\hbox{if }l\equiv 1\bmod 2 \cr
}\right\}\quad
\hbox{for } -1\gg l. \cr
}
$$
Below we consider the cases $m'=m$ only.
The vacuum vector in $V(\la)\o V(\la)^{\ast a}$
is $\bar{v} = \sum_{i} v_i \o v_i^*$ ($=\ket{vac}$)
where
$\{v_i\}$ is a base of $V(\la)$ and $\{v_i^*\}$ its dual base.
This $\bar{v}$ forms a one-dimensional submodule.
The subject of the present paper is
to calculate the vacuum expectation value of an arbitrary local operator.
A local operator $L$ is a linear operator
acting on a finite tensor product of~$V$:
$$
L\in \End(\underbrace{V \o\cdots\o V}_{n})
$$
where $V \o\cdots\o V$ is understood as $n$th to $1$st components
in $V^{\o\infty}$.
This acts on $V(\la)\o V(\la)^{\ast a}$ via the vertex operators.
The vacuum expectation value of $L$, or the correlation function,
is derived, in the same manner as in ref.\refno{\JMMN} for spin $1/2$, as
$$
\eqalignno{
\br{L}^{(\la)}_{z_n,\ldots,z_1}
&=
\ov{\bra{vac} \varrho^{(\la)}_{z_n,\ldots,z_1}(L)
\o \id_{V(\mu)^{\ast a}}\ket{vac}}
{\br{vac | vac}}
\cr
&=
\ov{\tr_{V(\la)}(q^{-2\rho}\varrho^{(\la)}_{z_n,\ldots,z_1}(L))}
{\tr_{V(\la)}(q^{-2\rho})}
&(4.2)\cr
}
$$
where $\rho = \La_0 + \La_1$ and
$$
\varrho^{(\la)}_{z_n,\ldots,z_1}(L)
=(\VO^{(n)}_{\la}(z_n,\ldots,z_1)^{-1}\circ (\id_{V(\la^{(n)})}\o L)
\circ \VO^{(n)}_{\la}(z_n,\ldots,z_1)),
$$
$$
\eqalignno{
\VO^{(n)}_{\la}&(z_n,\ldots,z_1)
\cr
=&(\VOt_{\la^{(n-1)}}^{\la^{(n)} V}(z_n)\o
\id_{\underbrace{\scriptstyle V \o\cdots\o V}_{n-1}})\circ\cdots\circ
(\VOt_{\la^{(1)}}^{\la^{(2)} V}(z_2)\o\id_{V})\circ
\VOt_{\la}^{\la^{(1)} V}(z_1),
\cr
\VO^{(n)}_{\la}&(z_n,\ldots,z_1)^{-1}
\cr
=&\VOt^{\la}_{\la^{(1)} V}(z_1)\circ
(\VOt^{\la^{(1)}}_{\la^{(2)} V}(z_2)\o\id_{V})
\circ\cdots\circ
(\VOt^{\la^{(n-1)}}_{\la^{(n)} V}(z_n)\o
\id_{\underbrace{\scriptstyle V \o\cdots\o V}_{n-1}})/
(g_{\la}g_{\la^{(1)}}\cdots g_{\la^{(n-1)}}).
\cr
}
$$
Let us set
$$
P_{i_n,\ldots,i_1}^{j_n,\ldots,j_1}(z_n,\ldots,z_1|\la)
=
\br{E_{i_n j_n}\o \cdots \o E_{i_1 j_1}}^{(\la)}_{z_n,\ldots,z_1}
$$
where $E_{ij} \cdot u_{j'} = \delta_{jj'} u_i$.

In the following we concentrate on calculations of
one-point functions
$$
P_i^j(z|\la)=\ov{1}{g_{\la}}
\ov{\tr_{V(\la)}\bigl(q^{-2\rho}
\VOt^{\la}_{\mu V, i}(z) \VOt_{\la, j}^{\mu V}(z) \bigr)}
{\tr_{V(\la)}\bigl(q^{-2\rho} \bigr)},
$$
$$
\la=(k-m)\La_0+m\La_1, \q \mu=m\La_0+(k-m)\La_1, \q 0\le m\le k
$$
for level $k=2$ (spin 1).
Note that $\rho = 2d + {1\over 2}\partial_\alpha$
in the present representations.

The specialized characters
$Z^{(\la)} = \tr_{V(\la)}$$ \bigl( q^{-2\rho} \bigr)$,
$\la=2\La_0, 2\La_1, \La_0+\La_1$,
which appeared as denominators,
are calculated straightforwardly.
The result is
$$
\eqalignno{
Z^{(2\La_i)}
&= q^{-(2\La_i,2\La_i)} \cdot (-q^2;q^2)_\infty (-q^4;q^4)_\infty,
&(4.3)\cr
Z^{(\La_0+\La_1)}
&= q^{-1} \cdot (-q^2;q^2)_\infty (-q^2;q^4)_\infty.
&(4.4)\cr
}
$$


Numerators of spin one-point functions are expressed by
vertex-operator two-point functions
$$
\eqalignno{
&
F^{(\la)}_{jk}(z_1/z_2)
= \tr_{V(\la)}\bigl( q^{-2\rho}
\VOt^{\la V}_{\sigma(\la), j}(z_1) \VOt^{\sigma(\la) V}_{\la, k}(z_2) \bigr).
\cr
}
$$
Let us first calculate these functions.
In ref.\refno{\JMMN}, they introduced some auxiliary boson
to simplify computations of traces slightly.
It is, however, not necessary to do so.
We perform direct calculations of traces.
Write
$$
d=\bar{d}^a+\bar{d}^\phi-{1\over 8}\partial_\alpha^2+{(\la,\la)\over 4}
$$
on $V(\la)$ where
$$
\bar{d}^a=-\sum_{m=1}^\infty mN_m^a, \q
\bar{d}^\phi=-\sum_{m={1\over 2},{3\over 2},\ldots} mN_m^\phi;
$$
$$
\eqalignno{
V_<(z_1,z_2,w_1,w_2)
&=\exp\bigl(-\sum_{m=1}^\infty a_{-m}{q^{m}\over[2m]}
  \xi_{m}(z_1,z_2,w_1,w_2) \bigr),
\cr
V_>(z_1,z_2,w_1,w_2)
&=\exp\bigl(\sum_{m=1}^\infty a_{m}{q^{m}\over[2m]}
  \xi_{-m}(z_1,z_2,w_1,w_2) \bigr),
\cr
\xi_m(z_1,z_2,w_1,w_2)&=w_1^m+w_2^m-q^{4m}z_1^m-q^{4m}z_2^m.
\cr
}
$$
When calculating traces directly,
we use the following formulas:
a trace on the boson Fock space
$$
\eqalignno{
\tr_{{\cal F}^a} &
\bigl( q^{-4\bar{d}^a}V_<(z_1,z_2,w_1,w_2)V_>(z_1,z_2,w_1,w_2)\bigr)
\cr
&=\ov{(q^6)_\infty^4}{(q^4)_\infty}
  (q^6{z_1\over z_2})_\infty(q^6{z_2\over z_1})_\infty \cdot
\ov{(q^6{w_1\over w_2})_\infty(q^6{w_2\over w_1})_\infty}
{\prod_{i,j=1,2}\{(q^2{w_i\over z_j})_\infty(q^{10}{z_j\over w_i})_\infty\}}
\cr
}
$$
where $(\cdot)_\infty = (\cdot;q^4)_\infty$
(see Appendix C for the normal ordering of boson exponentials);
a trace on the total NS fermion Fock space
$$
\eqalignno{
\tr_{{\cal F}^\phi} &
\bigl(\xi^{-2\bar{d}^\phi} :\phi(w_1)\phi(w_2): \bigr)\cr
&={(-\xi;\xi^2)_\infty \over q+q^{-1}} \cdot
  \bigl({w_1\over w_2}\bigr)^{1/2}
 \biggl[
  \sum_{N=0}^{+\infty}
  \bigl({w_1\over w_2}\bigr)^{N}
  \ov{q^{2N+1}+q^{-2N-1}}{1+\xi^{-2N-1}}
\cr
&\qq\qq\qq -
  \sum_{N=-\infty}^{-1}
  \bigl({w_1\over w_2}\bigr)^{N}
  \ov{q^{2N+1}+q^{-2N-1}}{1+\xi^{2N+1}}
 \biggr] \cr
&={(-\xi;\xi^2)_\infty \over q+q^{-1}} \cdot
  \bigl({w_1\over w_2}\bigr)^{1/2}\sum_{N\in\Z} f_N(\xi)\cdot
  \bigl({w_1\over w_2}\bigr)^N \cr
}
$$
where
$$
\eqalignno{
f_N(\xi)
&=\ov{q^{2N+1}+q^{-2N-1}}{1+\xi^{-2N-1}} \q\hbox{for } N\ge 0,
\cr
&=-\ov{q^{2N+1}+q^{-2N-1}}{1+\xi^{2N+1}} \q\hbox{for } N< 0;
\cr
}
$$
a trace on any one of the two sectors of the Ramond fermion Fock space
$$
\tr_{}
\bigl(\xi^{-2\bar{d}^\phi} :\phi(w_1)\phi(w_2): \bigr)
={(-q^4;q^4)_\infty \over q+q^{-1}} \cdot
 \biggl[
  {q^2{w_1\over w_2}\over 1-q^2{w_1\over w_2} }
 -
  {q^2{w_2\over w_1}\over 1-q^2{w_2\over w_1} }
 \biggr];
$$
and a trace on the total root lattice
$$
\tr_{\K[Q]}
  \bigl( \xi^{{1\over 4}\partial_\alpha^2} q^{-\partial_\alpha}
    y^{{1\over 2}\partial_\alpha}
  \bigr)
=\sum_{n\in\Z}
    \xi^{n^2} (q^{-2}y)^{n}
=\Theta_1(q^{-1}y^{{1\over 2}}|\xi)
$$
where $y={q^8z_1z_2\over w_1w_2}$.

For a technical reason,
when we treat with NS fermions,
we introduce a parameter~$\xi$ and calculate
$$
\eqalignno{
&
F^{(\la)}_{jk}(z_1,z_2|\xi)
= \tr_{V(\la)} \bigl(
q^{-2\rho'} \xi^{-2\bar{d}^\phi + {1\over 4}\partial_\alpha^2}
\VOt^{\la V}_{\sigma(\la), j}(z_1) \VOt^{\sigma(\la) V}_{\la, k}(z_2) \bigr)
\cr
}
$$
where
$\rho' = 2d' + {1\over 2}\partial_\alpha$ and
$d'=\bar{d}^a+{(\la,\la)\over 4}$.
Setting $\xi=q^2$ yields the desired function:
$F^{(\la)}_{jk}(z_1,z_2|q^2) = F^{(\la)}_{jk}(z_1,z_2)$.
Define a symmetrization/antisymmetrization operation
with respect to the parameter $\xi$
by
$$
\eqalignno{
f^{S}(\xi)&={1\over 2}(f(\xi)+f(-\xi)),\cr
f^{A}(\xi)&={1\over 2}(f(\xi)-f(-\xi)) \cr
}
$$
for an arbitrary function of $\xi$.
We sometimes denote $\sigma_0 = S$, $\sigma_1 = A$.

The results are listed below:
$$
\eqalignno{
F^{(2\La_i)}_{jk}(z|\xi)
&=q^{-4(2-k)}\cdot ( q^{-2}z^{-1} )^i \cdot b(z) \cdot
  \Res_{{w_1}} \Res_{{w_2}}
\biggl\{
\cr
&\q
  (w_1w_2)^{-{k\over 2}} \cdot A_k(w_1,w_2,z)
  \Bigl[ N(w_1,w_2,z|\xi) \Bigr]^{\sigma_i}
\biggr\},
\cr
&&(4.5a)\cr
}
$$
$$
\eqalignno{
F^{(\La_0+\La_1)}_{jk}(z)
&=-q^{4(k-1)}z^{{k\over 2}}\cdot b(z) \cdot
  \Res_{{w_1}} \Res_{{w_2}}
\biggl\{
  (w_1w_2)^{-1-{k\over 2}} \cdot A_k(w_1,w_2,z)
\cr
&\q \times
  (1-q^2)^2(-q^4;q^4)_\infty
  \Pi'(z^{-{k\over 2}}w_1,z^{-{k\over 2}}w_2)
  \Theta_1(({q^8z\over w_1w_2})^{{1\over 2}}|q^2)
\biggr\}
\cr
&&(4.6a)\cr
}
$$
for $(j,k)=(2,0),(0,2)$;
$$
\eqalignno{
F^{(2\La_i)}_{11}(z|\xi)
&=-q^{-4}\cdot ( q^{-2}z^{-1} )^i \cdot b(z) \cdot
  \Res_{{w_1}} \Res_{{w_2}}
\biggl\{
\cr
&\q
  {1\over w_1}  B(w_1,w_2,z)
  \Bigl[ M(w_1,w_2,z|\xi) \Bigr]^{\sigma_i}
\biggr\},
\cr
&&(4.5b)\cr
}
$$
$$
\eqalignno{
F^{(\La_0+\La_1)}_{11}(z)
=& q^{-1}(1-q^4)^2(-q^4;q^4)_\infty b(z)  \cr
&\quad\times  \Res_{{w_2}}
\biggl\{
  {1\over w_1} B(q^{-2}w_2,w_2,z)
  \Theta_1(({q^{10}z\over w_2^2})^{{1\over 2}}|q^2)
\biggr\}.
\cr
&&(4.6b)\cr
}
$$
Here are notations:
$$
\eqalignno{
A_0(w_1,w_2,z)=
&
\ov{(q^6{w_1\over w_2})_\infty (q^6{w_2\over w_1})_\infty }
{(q^{-2}w_1)_\infty (q^{-2}w_2)_\infty
(q^{-2}z^{-1}w_1)_\infty (q^{-2}z^{-1}w_2)_\infty}
\cr
&\qq\times
\ov{1}
{(q^{6}w_1^{-1})_\infty (q^{6}w_2^{-1})_\infty
(q^{10}zw_1^{-1})_\infty (q^{10}zw_2^{-1})_\infty }, \cr
&&(4.7a)\cr
\cr
A_2(w_1,w_2,z)=
&
\ov{(q^6{w_1\over w_2})_\infty (q^6{w_2\over w_1})_\infty }
{(q^{2}w_1)_\infty (q^{2}w_2)_\infty
(q^{-2}z^{-1}w_1)_\infty (q^{-2}z^{-1}w_2)_\infty}
\cr
&\qq\times
\ov{1}
{(q^{6}w_1^{-1})_\infty (q^{6}w_2^{-1})_\infty
(q^{6}zw_1^{-1})_\infty (q^{6}zw_2^{-1})_\infty }, \cr
&&(4.7b)\cr
\cr
B(w_1,w_2,z)=
&
\ov{(q^6{w_1\over w_2})_\infty (q^6{w_2\over w_1})_\infty }
{(q^{2}w_1)_\infty (q^{-2}w_2)_\infty
(q^{-2}z^{-1}w_1)_\infty (q^{-2}z^{-1}w_2)_\infty}
\cr
&\qq\times
\ov{1}
{(q^{6}w_1^{-1})_\infty (q^{6}w_2^{-1})_\infty
(q^{6}zw_1^{-1})_\infty (q^{6}zw_2^{-1})_\infty }; \cr
&&(4.7c)\cr
\cr
}
$$
$$
N(w_1,w_2,z|\xi)=
(1-q^2) (-\xi;\xi^2)_\infty \Pi({w_1\over w_2}|\xi)
  \Theta_1(({q^6z\over w_1w_2})^{1/2}|\xi),
$$
$$
M(w_1,w_2,z|\xi)=
(1-q^4) (-\xi;\xi^2)_\infty \varpi({w_1\over w_2}|\xi)
  \Theta_1(({q^6z\over w_1w_2})^{1/2}|\xi);
$$
$$
\eqalignno{
\Pi(w|\xi)
&= 1 - q\cdot f_0(\xi) \cr
&+ \sum_{m=1}^{+\infty} \bigl(w^m + w^{-m}\bigr)
  {1\over 2} \bigl[ (1-q^2)q^{-2m} +
      q \bigl(f_{m-1}(\xi) - f_{m}(\xi)\bigr) \bigr],
\cr
\varpi(w|\xi)
&=
(w-q^2)\sum_{n\in \Z} f_n(\xi) w^n \cr
&+ (q+q^{-1})
\bigl[
1 + (1-q^2) \sum_{n=1}^{+\infty}q^{-2n} w^n
\bigr],\cr
\Pi'(w_1,w_2)
&={1\over 2}\sum_{n=0}^{+\infty}
  \Bigl[
   w_1\Bigl( {w_1\over w_2} \Bigr)^n + w_2\Bigl( {w_2\over w_1} \Bigr)^n
  \Bigr] \cdot
  q^{-2n-2}(1+q^{4n+2});
\cr
}
$$
and
$$
b({z_1\over z_2}) =
{ (q^6;q^4)_\infty^4 \over (q^8;q^4)_\infty }
(q^6{z_1\over z_2};q^4)_\infty (q^2{z_2\over z_1};q^4)_\infty.
$$
We note that $\varpi(w|q^2)  =  q^{-1}  (1-q^4)  \delta(q^2w)$.

Spin one-point functions are
related to these functions.
Non zero ones are
$$
\eqalignno{
\br{E_{jj}}^{(\la_m)}_z
&=
P_{j}^{j}(z|\la_m)
=
\ov{c_{k-j}}{c_m} \cdot
\ov{1}{g_{\la_m} Z^{(\la_m)}}
\times
F^{(\la_m)}_{k-j,j}(q^{-2}|q^2).
\cr
}
$$
where $j,m = 0,1,2$ (for $k=2$).
Observe that they are independent of $z$.
{}From eq.(4.5) and eq.(4.6),
we have for the numerators
$$
\eqalignno{
F^{(2\La_i)}_{jk}(q^{-2}|\xi)
&=q^{-4(2-k)} \cdot b(q^{-2}) \cdot
  \Res_{{w_1}} \Res_{{w_2}}
\biggl\{
  (w_1w_2)^{-{k\over 2}} \cdot A_k(w_1,w_2,q^{-2})
\cr
&\q \times
  \Bigl[
    (1-q^2) (-\xi;\xi^2)_\infty \Pi({w_1\over w_2}|\xi)
     \Theta_1(({q^4\over w_1w_2})^{{1\over 2}}|\xi)
  \Bigr]^{\sigma_i}
\biggr\}
\cr
&&(4.8a)\cr
}
$$
and
$$
\eqalignno{
F^{(\La_0+\La_1)}_{jk}(q^{-2})
&=-q^{3k-4} \cdot b(q^{-2}) \cdot
  \Res_{{w_1}} \Res_{{w_2}}
\biggl\{
  (w_1w_2)^{-1-{k\over 2}} \cdot A_k(w_1,w_2,q^{-2})
\cr
&\q \times
  (1-q^2)(-q^4;q^4)_\infty
  \Pi'(q^kw_1,q^kw_2)
  \Theta_1(({q^6\over w_1w_2})^{{1\over 2}}|q^2)
\biggr\}
\cr
&&(4.9a)\cr
}
$$
for $(j,k)=(2,0),(0,2)$,
and
$$
\eqalignno{
F^{(2\La_i)}_{11}(q^{-2})
=& -{1\over 2} q^{-5}(1-q^4)^2(-q^2;q^4)_\infty b(q^{-2})\cdot
(q^4;q^4)_\infty (q^8;q^4)_\infty   \cr
&\quad\times  \Res_{{w_2}}
\biggl\{
  {1\over w_1}
  {1\over \bigl[ (q^{-2}w_2;q^2)_\infty (q^6w_2^{-1};q^2)_\infty \bigr]^2}
  \Theta_1(q^{3}w_2^{-1}|q^2)
\biggr\},
\cr
&&(4.8b)\cr
}
$$
$$
\eqalignno{
F^{(\La_0+\La_1)}_{11}(q^{-2})
=& q^{-1}(1-q^4)^2(-q^4;q^4)_\infty b(q^{-2})\cdot
(q^4;q^4)_\infty (q^8;q^4)_\infty   \cr
&\quad\times  \Res_{{w_2}}
\biggl\{
  {1\over w_1}
  {1\over \bigl[ (q^{-2}w_2;q^2)_\infty (q^6w_2^{-1};q^2)_\infty \bigr]^2}
  \Theta_1(q^{4}w_2^{-1}|q^2)
\biggr\}.
\cr
&&(4.9b)\cr
}
$$
We note that
$$
A_0(w_1,w_2,q^{-2})=
\ov{(q^6{w_1\over w_2};q^4)_\infty (q^6{w_2\over w_1};q^4)_\infty }
{\prod_{l=1,2} \{ (q^{-2}w_l;q^2)_\infty (q^6w_l^{-1};q^2)_\infty \}
},
$$
$$
A_2(w_1,w_2,q^{-2})=
\ov{(q^6{w_1\over w_2};q^4)_\infty (q^6{w_2\over w_1};q^4)_\infty }
{\prod_{l=1,2} \{ (w_l;q^2)_\infty (q^4w_l^{-1};q^2)_\infty \}
}.
$$
and that
$$
b(q^{-2})
=
{ (q^6;q^4)_\infty^4 \over (q^8;q^4)_\infty }
(q^4;q^4)_\infty (q^4;q^4)_\infty
=
(1-q^4)(q^4;q^4)_\infty(q^6;q^4)_\infty^4.
$$
These are our final expressions for spin one-point functions.

We can prove the following relations
which show `homogenuity' of the spin chain.

\proclaim Proposition 2.
We have relations
$$
P_j^j(z|2\La_0)=P_{2-j}^{2-j}(z|2\La_1),\q j=0,1,2; \qq
P_0^0(z|\La_0+\La_1)=P_{2}^{2}(z|\La_0+\La_1).
$$
Or equivalently, in terms of the function $F$ they are written as
$$
\eqalignno{
F^{(2\La_0)}_{2-j,j}(q^{-2}|q^2)
  &= q^{2j-2}\cdot
     F^{(2\La_1)}_{j,2-j}(q^{-2}|q^2),\q j=0,1,2; \cr
F^{(\La_0+\La_1)}_{20}(q^{-2})
  &= q^{-2}\cdot
     F^{(\La_0+\La_1)}_{02}(q^{-2}), \cr
}
$$
respectively.
\par

\Proof
Noticing a relation
$\Theta_1^{\sigma_i}(q^2z|q^2) = q^{-2}z^{-2}
\Theta_1^{\sigma_{1-i}}(z|q^2) $ ($i=0,1$, $\sigma_0 = S$, $\sigma_1 = A$),
and
$$
A_2(q^{-2}w_1,q^{-2}w_2,q^{-2}) = A_0(w_1,w_2,q^{-2}),
$$
we can prove it easily
(by changing integral variables, e.g., $q^2w_j$ by $w_j$, etc.)
\hfill\qed

Here we give a list of expansions in $q$ of our integral representations.
$$
\eqalignno{
P_2^2(2\La_0)
 &= 1-2q^2+5q^6-2q^8-9q^{10}+3q^{12}+19q^{14}+O(q^{16}); \cr
P_0^0(2\La_0)
 &= 2q^4+q^6-4q^8-9q^{10}+7q^{12}+19q^{14}+O(q^{16}); \cr
P_1^1(2\La_0)
 &= 2q^2-2q^4-6q^6+6q^8+18q^{10}-10q^{12}-38q^{14}+O(q^{16}); \cr
}
$$
$$
\eqalignno{
P_0^0(\La_0+\La_1)
 &= 2q^2-4q^4+2q^6+6q^{10}-4q^{12}-6q^{14}+O(q^{16}); \cr
P_1^1(\La_0+\La_1)
 &= 1-4q^2+8q^4-4q^6-12q^{10}+8q^{12}+12q^{14}+O(q^{16}). \cr
}
$$
{}From this list we get
$$
\br{s^z}^{(2\La_1)} = 1 - 2q^2 - 2q^4 + 4q^6 + 2q^8 - 4q^{12} + O(q^{16})
\biggl( = {(q^2;q^2)_\infty^2 \over (-q^4;q^4)_\infty }\biggr)
$$
where $s^z=\hbox{diag}(1,0,-1)$,
which agrees with the known Bethe Ansatz result\refhat{\DJMO}
(see Appendix B).
We also note that
the expectation value with respect to the vacuum
in $V(\La_0+\La_1)\o V(\La_0+\La_1)^{\ast a}$
is zero, since from Proposition 2
$$
\eqalignno{&
\br{s^z}^{(\La_0+\La_1)}
= P^0_0(z|\La_0+\La_1)-P^2_2(z|\La_0+\La_1)
= 0.
\cr
}
$$
This again agrees with the result in ref.\refno{\DJMO}.



Finally,
as was mentioned in section 3
we have not justified rigorously
the correctness of our vertex operators.
The proof remains to be completed.

\beginsection Acknowledgements

The auther wishes to thank
Professors
Tetsuji Miwa, Michio Jimbo, and Tetsuji Tokihiro
for many helpful discussions and suggestions.

\beginsection Appendix A. The Hamiltonian

Let us define symmetric and anti-symmetric parts of $h$
(defined by eq.(4.1))
by
$$
\eqalignno{
h_S &= {1\over 2}(h+P h P), \cr
h_A &= {1\over 2}(h-P h P) \cr
}
$$
where  $P$ is the permutation operator in $\End( V^{(k)}\o V^{(k)} )$
such that $P(a\o b)=b\o a$.

\def\ch{{\rm ch}}
\def\sh{{\rm sh}}

An explicit form of the spin-1 analog of the XXZ Hamiltonian
(thus, $k=2$) is given as follows.
We set $q=-\e^{-\la}$ below ($0<\la<+\infty$).
$$
\eqalignno{
h_S=&
 s^1\o s^1 +  s^2\o s^2 +  \ch(2\la) \cdot s^3\o s^3 -
 \Bigl(\sum_{j=1}^3 s^j\o s^j \Bigr)^2 \cr
&+ 2\sh^2(\la) \cdot
 \Bigl[ (s^3)^2\o \id + \id\o (s^3)^2 -
 (s^3\o s^3)^2 - 2\cdot\id\o\id  \Bigr] \cr
&+ (2+4\e^{-\la}\ch^2(\la)) \cdot (s^1\o s^1 + s^2\o s^2) s^3\o s^3 \cr
&+ (2+\e^{\la}) \cdot  s^3\o s^3 (s^1\o s^1 + s^2\o s^2), \cr
h_A=&
-\sh(2\la) ( s^3\o \id - \id\o s^3  ). \cr
}
$$
We see that
the anti-symmetric part has the form $s^3\o \id - \id\o s^3$.
This implies that the summation of $h_A$ over all sites  vanishes.
Thus we conclude that the Hamiltonian has the form
$$
\eqalignno{
&
H
=\sum_{l\in\Z}
 \cdots \o \head{1}^{l+2}\o \head{{h_S}}^{{l+1\q l}} \o
 \head{1}^{l-1}\o \cdots.
\cr
}
$$

Let us see the $q\to 0$ limit of the Hamiltonian.
In the limit $q\to 0^-$ ($\la \to +\infty$) the Hamiltonian
tends to an antiferromagnetic Ising Hamiltonian
with appropriate re-normalization;
thus we say that the limit~$q\to 0$ is an Ising limit.
Define
$$
h^{Ising}
= \lim_{\la\to +\infty} {1\over 2}{1\over \sh k\la} h
= - \lim_{q\to 0^-} {(-1)^k\over q^k-q^{-k}} h,
$$
or more directly
$$
h^{Ising}
= - \lim_{q\to 0^-} \biggl[{d\over dz}\Rc(z,1)\biggr]_{z=1}.
$$
Then we find $h^{Ising} \cdot u_i\o u_j = H(i,j) u_i\o u_j$,
$i,j=0,1,\ldots,k$,
where $H(i,j) = -j$ for $i+j \le k$, $=i-k$ for $i+j \ge k$.
Especially its symmetric part is
$h_S^{Ising} \cdot u_i\o u_j = H_S(i,j) u_i\o u_j$
where $H_S(i,j) = -{1\over 2}(i+j)$ for $i+j \le k$,
$={1\over 2}(i+j-2k)$ for $i+j \ge k$.
Explicit forms are as follows (for $k=2$):
$$
\eqalignno{
h_S^{Ising}=&
 {1\over 2} \Bigl[ s^3\o s^3 + (s^3)^2\o \id + \id\o (s^3)^2 -
 (s^3\o s^3)^2 - 2\cdot\id\o\id  \Bigr], \cr
h_A^{Ising}=&
-{1\over 2} ( s^3\o \id - \id\o s^3  ). \cr
}
$$


\beginsection Appendix B. The result of Ref.\refno{\DJMO}

In ref.\refno{\DJMO},
an exact expression for the one-point function $\br{s^z}^{(\la_m)}$
(in vertex model language, it is called the mean staggered polarization)
is obtained for $m= 0,1, \ldots, k$ and for arbitrary spin~$k/2$
using the Bethe Ansatz.
Their result is (we change notations slightly)
$$
\br{s^z}^{(\la_m)}
= \prod_{n=1}^{+\infty}
  \Bigl( {1+q^{(k+2)n}\over 1-q^{(k+2)n}}\Bigr)^2 \cdot
  \bigl[ m-{k\over 2}-2\sum_{l=0}^{+\infty}(-)^l
    \bigl( f(l,k-m)-f(l,m) \bigr)
  \bigr]
$$
$m=0,1,\ldots,k$, where $k/2$ is the spin
($k$ is the level in the present paper)
and
$$
f(l,m)=[(k+2)l+m+1]\cdot {q^{2[(k+2)l+m+1]}\over 1-q^{2[(k+2)l+m+1]}}.
$$
It gives series expansions
of $\br{s^z}^{(k\La_1)}$ in~$q$:
$$
\eqalignno{
k=& 1:
  {\scriptstyle {1\over 2}}-2q^2+2q^4+2q^8-4q^{10}+
  2q^{16}-2q^{18}+4q^{20}+O(q^{22}); \cr
k=& 2:
  1-2q^2-2q^4+4q^6+2q^8-4q^{12}+2q^{16}-6q^{18}+O(q^{22}); \cr
}
$$
and so on.

We found their infinite-product representations.
Here we report the results.
$$
\eqalignno{
\br{s^z}^{(\La_1)}
&={1\over 2}\br{\sigma^z}^{(\La_1)}
={1\over 2}\cdot {(q^2;q^2)_\infty^2 \over (-q^2;q^2)_\infty^2 }
\q\hbox{for $k=1$};\cr
\br{s^z}^{(2\La_1)}
&={(q^2;q^2)_\infty^2 \over (-q^4;q^4)_\infty }
={(q^4;q^4)_\infty^2 \over (-q^2;q^2)_\infty^2 (-q^4;q^4)_\infty }
\q\hbox{for $k=2$}.\cr
}
$$
(We checked them up to $q^{150}$.)

\beginsection Appendix C. Normal ordering of exponentials of boson

We collect here formulas for the normal ordering of exponentials of
boson operators.
We recall that
$$
\eqalignno{
\VO_2(z)&=
F_<(z)F_>(z)
\e^{\alpha} (-q^4z)^{{1\over 2}\partial_\alpha},
\cr
x^+(z)&=
\eta
E_<^+(z)E_>^+(z)
\phi(z)
\e^\alpha z^{{1\over 2}+{1\over 2}\partial_\alpha},
\cr
x^{-}(z)&=
\eta
E_<^-(z)E_>^-(z)
\phi(z)
\e^{-\alpha} z^{{1\over 2}-{1\over 2}\partial_\alpha},
\cr
}
$$
where
$$
\matrix{
{\displaystyle
F_<(z)=
\exp\bigl(\sum_{m=1}^\infty{a_{-m}\over[2m]}q^{5m}z^m\bigr)},&
{\displaystyle
F_>(z)=
\exp\bigl(-\sum_{m=1}^\infty{a_{m}\over[2m]}q^{-3m}z^{-m}\bigr)},\cr
{\displaystyle
E_<^+(z)=
\exp\bigl(\sum_{m=1}^\infty{a_{-m}\over[2m]}q^{-m}z^m\bigr)},&
{\displaystyle
E_>^+(z)=
\exp\bigl(-\sum_{m=1}^\infty{a_{m}\over[2m]}q^{-m}z^{-m}\bigr)},\cr
{\displaystyle
E_<^-(z)=
\exp\bigl(-\sum_{m=1}^\infty{a_{-m}\over[2m]}q^{m}z^m\bigr)},&
{\displaystyle
E_>^-(z)=
\exp\bigl(\sum_{m=1}^\infty{a_{m}\over[2m]}q^{m}z^{-m}\bigr)},\cr
{\displaystyle
\eta=
(q+q^{-1})^{1/2}}, &
{\displaystyle
[a_{m},a_{-n}]=\delta_{m,n}{[2m]^2\over m} }.\cr
}
$$
We have the following:
$$
\eqalignno{
F_>(z_1)F_<(z_2)&=(1-q^{2}z_2/z_1)F_<(z_2)F_>(z_1), \cr
F_>(z_1)E^-_<(z_2)&=\ov{1}{1-q^{-2}z_2/z_1}E^-_<(z_2)F_>(z_1), \cr
E^-_>(z_1)F_<(z_2)&=\ov{1}{1-q^{6}z_2/z_1}F_<(z_2)E^-_>(z_1), \cr
E^-_>(z_1)E^-_<(z_2)&=(1-q^{2}z_2/z_1)E^-_<(z_2)E^-_>(z_1), \cr
&\cr
F_>(z_1)E^+_<(z_2)&=(1-q^{-4}z_2/z_1)E^+_<(z_2)F_>(z_1), \cr
E^+_>(z_1)F_<(z_2)&=(1-q^{4}z_2/z_1)F_<(z_2)E^+_>(z_1), \cr
E^+_>(z_1)E^+_<(z_2)&=(1-q^{-2}z_2/z_1)E^+_<(z_2)E^+_>(z_1), \cr
&\cr
E^+_>(z_1)E^-_<(z_2)&=\ov{1}{1-z_2/z_1}E^-_<(z_2)E^+_>(z_1), \cr
E^-_>(z_1)E^+_<(z_2)&=\ov{1}{1-z_2/z_1}E^+_<(z_2)E^-_>(z_1). \cr
}
$$

\vfill\eject

\beginsection References \par

\item{\DFJMN)}
B. Davies, O. Foda, M. Jimbo, T. Miwa and A. Nakayashiki:
Commun. Math. Phys. {\bf 151} (1993) 89.
\par

\item{\JMMN)}
M. Jimbo, K. Miki, T. Miwa and A. Nakayashiki:
Phys. Lett. {\bf A168} (1992) 256.
\par

\item{\IIJMNT)}
M. Idzumi, T. Tokihiro, K. Iohara, M. Jimbo, T. Miwa and T. Nakashima:
Int. J. Mod. Phys. {\bf A 8} (1993) 1479.
\par

\item{\FrJ)}
I.B. Frenkel and N.H. Jing:
Proc. Nat'l. Acad. Sci. USA {\bf 85} (1988) 9373.
\par

\item{\CP)}
V. Chari and A. Pressley:
Commun. Math. Phys. {\bf 142} (1991) 261.
\par

\item{\LepowskyP)}
J. Lepowsky and M. Primc:
Lecture Notes in Mathematics {\bf 1052} (1984) 194.
\par

\item{\Bernard)}
D. Bernard:
Lett. Math. Phys. {\bf 17} (1989) 239.
\par

\item{\TsuchiyaK)}
A. Tsuchiya and Y. Kanie:
Adv. Stud. Pure Math. {\bf 16} (1988) 297.
\par

\item{\FR)}
I.B. Frenkel and N.Yu. Reshetikhin:
Commun. Math. Phys. {\bf 146} (1992) 1.
\par

\item{\DJO)}
E. Date, M. Jimbo and M. Okado:
Osaka Univ. Math. Sci. preprint {\bf 1} (1991).
\par

\item{\Ka2)}
M. Kashiwara:
RIMS preprint {\bf 756} (1991).
\par

\item{\DJMO)}
E. Date, M. Jimbo, K. Miki and M. Okado:
Int. J. Mod. Phys. {\bf A 7},  Suppl. 1A (1992) 151.
\par

\end